\documentclass[showpacs,dvipdfm,aps,prb,amsmath,amssymb,superscriptaddress,notitlepage,twocolumn]{revtex4-1}
\usepackage[dvips]{graphicx}% Include figure files
\usepackage{epstopdf}
\usepackage{dcolumn}% Align table columns on decimal point
\usepackage{bm}% bold math
\usepackage{pslatex}
\usepackage{multirow} 
\usepackage{array}
\usepackage[T1]{fontenc}
\usepackage[utf8]{inputenc}
\usepackage{dcolumn}% Align table columns on decimal point
\usepackage[sort&compress]{natbib}
\newcommand{\fref}[1]{Fig.~\ref{#1}}
\renewcommand{\eqref}[1]{Eq.~(\ref{#1})}
\newcommand{\eref}[1]{~(\ref{#1})}
\renewcommand{\deg}{^{\circ}}
\newcommand{\cso}{SeCuO$_3$ }
\newcommand{\csos}{SeCuO$_3$}
\renewcommand{\bm}[1]{\mathbf{#1}}

\usepackage[pdfauthor={Mirta Herak},pdftitle={Magnetic anisotropy of CuSeO3},colorlinks=true,citecolor=blue,linkcolor=blue,urlcolor=blue,bookmarks=true]{hyperref}
\begin{document}

%\preprint{APS/123-QED}

\title{\texorpdfstring{Magnetic anisotropy of spin tetramer system \cso studied by torque magnetometry and ESR spectroscopy}{Magnetic anisotropy of spin tetramer system SeCuO3 studied by torque magnetometry and ESR spectroscopy}}% Force line breaks with \\

\author{Mirta~Herak}\email{mirta@ifs.hr}
\affiliation{Institute of Physics, Bijeni\v{c}ka c. 46, HR-10000 Zagreb, Croatia}
\author{Antonija~Grubi\v{s}i\'{c}~\v{C}abo}
\affiliation{Institute of Physics, Bijeni\v{c}ka c. 46, HR-10000 Zagreb, Croatia}
\author{Dijana~\v{Z}ili\'{c}}
\affiliation{Ru\dj{}er Bo\v{s}kovi\'{c} Institute,  Bijeni\v{c}ka c. 54, HR-10000, Zagreb, Croatia}
\author{Boris~Rakvin}
\affiliation{Ru\dj{}er Bo\v{s}kovi\'{c} Institute,  Bijeni\v{c}ka c. 54, HR-10000, Zagreb, Croatia}
\author{Kre\v{s}imir~Salamon}
\affiliation{Institute of Physics, Bijeni\v{c}ka c. 46, HR-10000 Zagreb, Croatia}
\author{Ognjen~Milat}
\affiliation{Institute of Physics, Bijeni\v{c}ka c. 46, HR-10000 Zagreb, Croatia}
\author{Helmuth~Berger}
\affiliation{Institute of Physics of Complex Matter, EPFL, 1015 Lausanne, Switzerland}
\date{\today}

\begin{abstract}
We present an experimental study of macroscopic and microscopic magnetic anisotropy of a spin tetramer system \cso using torque magnetometry and ESR spectroscopy. Large rotation of macroscopic magnetic axes with temperature observed from torque magnetometry agrees reasonably well with the rotation of the $\bm{g}$ tensor above $T \gtrsim 50$~K. Below 50~K, the $\bm{g}$ tensor is temperature independent, while macroscopic magnetic axes continue to rotate. Additionally, the susceptibility anisotropy has a temperature dependence which cannot be reconciled with the isotropic Heisenberg model of interactions between spins. ESR linewidth analysis shows that anisotropic exchange interaction must be present in \csos. These findings strongly support the presence of anisotropic exchange interactions in the Hamiltonian of the studied system. Below $T_N=8$~K, the system enters a long - range antiferromagnetically ordered state with easy axis along the $<\bar{1} 0 1>^*$ direction. Small but significant rotation of magnetic axes is also observed in the antiferromagnetically ordered state suggesting strong spin-lattice coupling in this system. 
\end{abstract}

\maketitle
%
%%%%%%%%%%%%%%%%%%%%%%%%%%%%%%%%%%%%%%%%%%%%%%%%%%%%%%%%%%%%%%%%%%%%%%%%%%%%%%%%%%%
%
%                               INTRODUCTION
%
%%%%%%%%%%%%%%%%%%%%%%%%%%%%%%%%%%%%%%%%%%%%%%%%%%%%%%%%%%%%%%%%%%%%%%%%%%%%%%%%%%%%
%
%
\section{Introduction}\label{sec:intro}
\begin{figure}[tb]
\centering
\includegraphics[clip,width=\columnwidth]{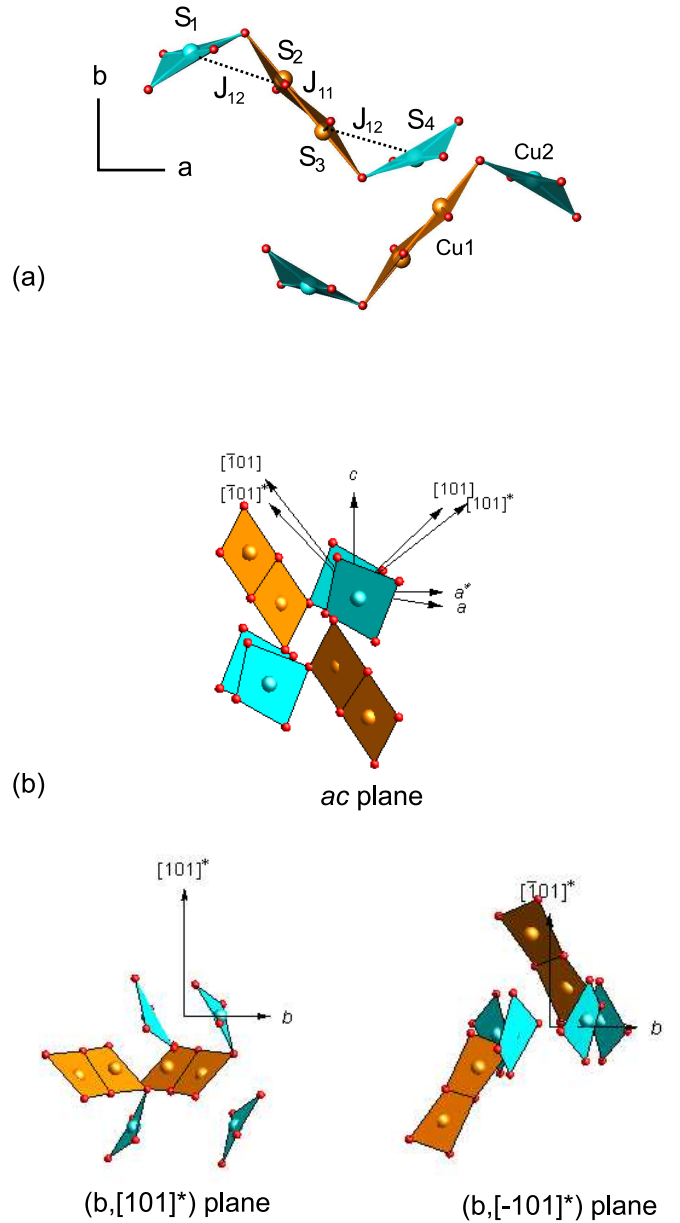}
	\caption{(Color online.) (a) Two magnetically inequivalent isolated spin tetramers in \csos. Orange (dark) and cyan (light) squares represent crystallographically inequivalent Cu1O$_4$ and Cu2O$_4$ irregular squares. Small red spheres represent oxygen atoms. Se atoms are not shown for the sake of clarity. $J_{11}$ and $J_{12}$ represent two different intratetraherdal interaction paths. (b) Unit cell of \cso with three crystallographic planes in which the measurements were performed.}
	\label{fig1}
\end{figure}

\indent Low-dimensional spin systems present a fertile ground for the study of magnetism. Their relatively simple magnetic lattices are ideal for analytical and numerical theoretical investigations. At the same time, such systems can be found in real materials thus enabling a comparison of experimental study with theory. Of special interest are spin $S=1/2$ systems in which the interactions of spins are usually well described by isotropic Heisenberg Hamiltonian and the magnetic anisotropy mostly comes from the $g$ factor anisotropy. Presence of small anisotropic exchange interaction can have a profound influence on the ground state and low-energy excitations of these systems. For example, presence of staggered $\bm{g}$ tensor and/or Dzyaloshinskii-Moriya interaction\cite{Dzyaloshinsky-1958,Moriya-1960,*Moriya-1960PR} (DMI) in $S=1/2$ one-dimensional Heisenberg antiferromagnet (1D HAF) opens a gap in the excitation spectrum in finite magnetic field.\cite{Dender-1997,Oshikawa-1997,Affleck-1999,*Affleck-2000} \\
\indent Influence of the small exchange anisotropy on the bulk magnetic susceptibility of Heisenberg antiferromagnets has received relatively little attention because bulk methods are often not sensitive enough in comparison to local probes such as electron spin resonance (ESR) and nuclear magnetic resonance (NMR). Nevertheless, there are few examples where the DMI significantly influences bulk susceptibility of low-dimensional Heisenberg antiferromagnets. In 1D HAF DMI induces anisotropic susceptibility enhancement at temperatures $T <J/k_B$, where $J$ is the intrachain interaction.\cite{Affleck-1999,*Affleck-2000, Feyerherm-2000, Umegaki-2009} Another example is magnetic susceptibility anisotropy of the kagome antiferromagnet with the DMI and symmetric anisotropic exchange which was recently studied by the numerical linked cluster method and exact diagonalization.\cite{Rigol-2007PRL, Rigol-2007} This study showed that the presence of the DMI can either suppress or enhance in and out of plane susceptibilities which has significant influence in the temperature dependence of the susceptibility anisotropy.\cite{Han-2012}\\
\indent Low-field torque magnetometry is a powerful method for measurement of macroscopic (bulk) magnetic anisotropy. In a paramagnetic substance, a measured component of torque is proportional to the magnetic susceptibility anisotropy $\Delta \chi$, i.e., the difference between maximal and minimal components of the susceptibility tensor in the plane of measurement. Direct measurement of $\Delta \chi$ in HAFs is obviously much more sensitive to the detection of the susceptibility enhancements or suppressions emerging from small anisotropy of exchange than the standard magnetic susceptibility measurement.\cite{HerakPRB-2011} Another advantage of torque magnetometry over bulk susceptibility measurement is a routine detection of the rotation of macroscopic magnetic axes with temperature. However, change of the $\bm{g}$ tensor with temperature also influences susceptibility anisotropy and is reflected in torque. In order to distinguish contributions of the exchange anisotropy from the $g$-factor anisotropy in torque measurements, it is necessary to combine torque magnetometry with the ESR measurements. In this work, we use this methodology to study the magnetic anisotropy of a spin tetramer system \csos.\\
\indent \cso crystallizes in monoclinic unit cell with space group $P2_1/n$.\cite{Effenberger-1986} There are two crystallographically inequivalent copper atoms, Cu1 and Cu2. Each Cu is surrounded by six oxygen atoms which form very distorted elongated CuO$_6$ octahedra. In both octahedra, four nearest equatorial oxygens form distored CuO$_4$ squares (see \fref{fig1}). Two apical oxygens are at distances $>2.36$~\AA~ so we expect only the four nearest oxygens forming a CuO$_4$ distorted square to take part in the Cu-O-Cu type of superexchange. When this is taken into account the magnetic lattice of \cso is formed of isolated spin tetramers [\fref{fig1}(a)].\cite{Zivkovic-2012} Cu1 ions form dimers by edge sharing of two oxygen atoms from coordination square (orange CuO$_4$ squares in \fref{fig1}). Each Cu2 shares one oxygen with one Cu1. From the inspection of Cu -- Cu distances and Cu -- O -- Cu angles two different antiferromagnetic exchange energies are expected in tetramer. Keeping notation of Ref. \onlinecite{Zivkovic-2012} we term superexchange between Cu1 and Cu1 $J_{11}$ and between Cu1 and Cu2 $J_{12}$ [see \fref{fig1}(a)]. The proposed Hamiltonian of this system is then
\begin{equation}\label{eq:Hamiltonian}
	\mathcal{H} = J_{12}(\bm{S}_1 \cdot \bm{S}_2 + \bm{S}_3\cdot \bm{S}_4) + J_{11} \bm{S}_2 \cdot \bm{S}_3.
\end{equation}
While magnetic susceptibility measurements give large intratetrahedral interactions $J/k_B\sim 200$~K, \cso orders antiferromagnetically at $T\approx 8$~K, suggesting that weak intertetrahedron interaction ($J_{inter}\ll J_{11(12)}$) is also present.\cite{Zivkovic-2012} It has been shown\cite{Zivkovic-2012} that the magnetic susceptibility of \cso could be described by the spin tetramer model\cite{Emori-1975} only down to $\sim 90$~K. Below that temperature, additional interaction was suggested to appear in Hamiltonian \eref{eq:Hamiltonian}, but the real reason for the deviation of the susceptibility from the tetramer model remains unclear. ESR measurements along two different directions revealed temperature change of $g$ factor, and torque measurements revealed a large rotation of macroscopic magnetic axes with temperature.\cite{Zivkovic-2012} In this work we investigate the connection between the temperature dependence of the $\bm{g}$ tensor and the rotation of macroscopic magnetic axes which will allow us to draw some conclusions about the interactions in this system.\\
\indent We present this work as follows. In Sec. \ref{sec:exp} we briefly describe the experimental methods we used. In Sec. \ref{sec:resul} we present results of torque magnetometry (Sec. \ref{sec:aniso}) and ESR spectroscopy (Sec. \ref{sec:esr}). We discuss the results in Sec. \ref{sec:disc} and summarize conclusions in Sec. \ref{sec:concl}.
%
%%%%%%%%%%%%%%%%%%%%%%%%%%%%%%%%%%%%%%%%%%%%%%%%%%%%%%%%%%%%%%%%%%%%%%%%%%%%%%%%%%%%%%%%%%%%%
%
%                            Experimental
%
%%%%%%%%%%%%%%%%%%%%%%%%%%%%%%%%%%%%%%%%%%%%%%%%%%%%%%%%%%%%%%%%%%%%%%%%%%%%%%%%%%%%%%%%%%%%%
\section{Experimental}\label{sec:exp}
\indent The single crystal of SeCuO$_3$ used in all measurements has been grown by a standard
chemical vapor phase method as previously described.\cite{Zivkovic-2012} The mass of the sample was $1.03\pm 0.03$~mg.\\
\indent The orientation of the crystal axes with respect to the sample morphology was determined using a Laue camera in back-scattering geometry. Several spectra, taken with the $b$ axis parallel or perpendicular to the x-ray beam, were subsequently simulated using ORIENTEXPRESS 3.3 software\cite{OExpress} and the axes were unambiguously determined. The high quality of the crystal was ascertained by the sharp Laue reflection and the X-band ESR measurements.\\
\indent Magnetic torque was measured with a home-built torque apparatus which uses torsion of thin quartz fibre for torque measurement. The sample holder is made of ultra pure quartz and has an absolute resolution of $10^{-4}$~dyn~cm. Measurements were made in fields up to $8$~kOe in the temperature range 2 - 330~K. Three different crystallographic planes were probed: the $ac$ plane, the plane spanned by axis $b$ and $b \times [1 0 1] = [1 0 \bar{1}]^*$ axis and plane spanned by $b$ axis and $b \times [\bar{1} 0 1] = [1 0 1]^*$ axis. Throughout this paper, we term the second one the $(b,\;[1 0 \bar{1}]^*)$ plane and the third one the $(b, [1 0 1]^*)$ plane. These three planes of measurement are shown in \fref{fig1}(b) with all significant axes mentioned in the text. The precision of the torque goniometer was $0.2\deg$.\\
\indent X-band ESR measurements were performed in the temperature range from 8~K to room temperature using Bruker Elexsys 580 FT/CW X-band spectrometer equipped with a standard Oxford Instruments model DTC2 temperature controller. The microwave frequency was $\sim9.7$~GHz with the magnetic field modulation amplitude of 5~G at 100~kHz. At 8~K the signal disappeared due to the establishment of long-range antiferromagnetic (AFM) order observed previously.\cite{Zivkovic-2012} ESR spectra were recorded at $5\deg$ steps and the rotation was controlled by a home-made goniometer with the accuracy of $1\deg$. In both torque and ESR measurements the uncertainty of $\approx 3\deg$ was related to the optimal deposition of the crystal on the quartz sample holder.
%
%
%
%%%%%%%%%%%%%%%%%%%%%%%%%%%%%%%%%%%%%%%%%%%%%%%%%%%%%%%%%%%%%%%%%%%%%%%%%%%%%%%%%%%%%%%%%%
%
%                              RESULTS
%
%%%%%%%%%%%%%%%%%%%%%%%%%%%%%%%%%%%%%%%%%%%%%%%%%%%%%%%%%%%%%%%%%%%%%%%%%%%%%%%%%%%%%%%%%%
%
%
\section{Results}\label{sec:resul}
%
%%%%%%%%%%%%%%%%%%%%%%%%%%%%%%%%%%%%%%%%%%%%%%%%%%%%%%%%%%%%%%%%%%%%%%%%%%%%%%%%%%%%%%%%%%
%
%                   TORQUE MAGNETOMETRY
%\bar{}
%%%%%%%%%%%%%%%%%%%%%%%%%%%%%%%%%%%%%%%%%%%%%%%%%%%%%%%%%%%%%%%%%%%%%%%%%%%%%%%%%%%%%%%%%%
%
\subsection{Torque magnetometry}\label{sec:aniso}
\indent A sample with finite magnetization $\mathbf{M}$ (induced or spontaneous) in finite magnetic field $\mathbf{H}$ will experience magnetic torque $\boldsymbol{\tau}$
\begin{equation}\label{eq:torque}
	\boldsymbol{\tau} = V\: \mathbf{M} \times \mathbf{H},
\end{equation}
where $V$ is the volume of the sample. In case of linear response of magnetization to magnetic field we have
\begin{equation}\label{eq:inducedM}
	\mathbf{M} = \hat{\chi}\: \mathbf{H}
\end{equation}
where $\hat{\chi}$ is the susceptibility tensor. In experiment, the magnetic field usually rotates in one plane, e.g., the $xy$ plane, and only the component of torque $\tau_z$ perpendicular to that plane is measured. Combining \eqref{eq:inducedM} and \eqref{eq:torque} we obtain for measured torque component $\tau_z$
\begin{equation}\label{eq:angularTorque}
	\tau_z = \dfrac{m}{2M_{mol}}\: H^2\: \Delta \chi_{xy} \: \sin(2\phi-2\phi_0)
\end{equation}
where $m$ is the mass of the sample, $M_{mol}$ is the molar mass, $H$ is the applied field, $\phi$ is the goniometer angle, and $\phi_0$ is the angle at which the magnetic field is parallel to the $x$ axis. Phase $\phi_0$ is introduced only because goniometer angle $\phi=0$ does not generally correspond to the direction of the axis $x$ of the sample. $\Delta \chi_{xy} = \chi_x - \chi_y$ is the magnetic susceptibility anisotropy in the $xy$ plane measured in emu/mol.\footnote{$\chi_{x(y)}$ is the magnetic susceptibility along the $x(y)$ axis. $x$ and $y$ axes can be principal magnetic axes of the system, but more generally these are directions of maximal and minimal susceptibility, $\chi_{max}$ and $\chi_{min}$, in the $xy$ plane.} When \eqref{eq:angularTorque} applies and $\phi_0$ does not change with temperature, we obtain temperature dependence of susceptibility anisotropy by measuring temperature dependence of torque. This is achieved by applying constant magnetic field in the direction of maximal or minimal torque. In the case when $\phi_0$ changes with temperature, it is possible to measure temperature dependence of both $\phi_0$ and the total susceptibility anisotropy $\chi_{max}-\chi_{min}$. The direction of $\chi_{max}$ coincides with $\phi_0$.
At $\phi_0$, the torque is zero which means that the magnetization is parallel to the magnetic field, so $\phi_0$ represents the direction of one of the magnetic axes. For simplicity, we consider here only a simple case in which the system has two magnetic axes in the plane of measurement and the third axis is perpendicular to them. This applies to our case. Change of $\phi_0$ with temperature represents a rotation of the magnetic axes in the plane of measurement, i.e., a change of direction of $\chi_{max}$ and $\chi_{min}$ in that plane.\\
\indent At this point, we should establish the connection between magnetic torque and the $g$-factor anisotropy measured by ESR since our interpretation of the measured data is based on that connection. From \eqref{eq:angularTorque} we see that measured $\tau_z \propto \chi_x - \chi_y$. For a paramagnet in low field $\chi_x \propto g_x^2$, which leads to $\tau_z \propto g_x^2-g_y^2$. For isotropic Heisenberg spin $S=1/2$ systems, the $g$-factor anisotropy is usually the dominant contribution to the susceptibility anisotropy and the contribution of small anisotropic exchange can be neglected in comparison, especially at temperatures $T>J/k_B$. This applies well to the Cu$^{2+}$ spin $S=1/2$ systems, such as \csos. Directions $x$ and $y$ at which the torque is zero are directions of maximal and minimal components of $\bm{g}$ tensor in the $xy$ plane. The rotation of the $\bm{g}$ tensor is thus reflected directly in the change of phase $\phi_0$ of measured torque [see \eqref{eq:angularTorque}].\\
\indent It is important to emphasize that \eqref{eq:angularTorque} is valid when \eref{eq:inducedM} applies for the paramagnetic state, but also for the collinear antiferromagnetic state in small applied fields. In collinear antiferromagnetically ordered systems, angular dependence of torque will generally deviate from \eqref{eq:angularTorque} if magnetic field is large enough to induce reorientation of spins.\cite{Herak-2013, Herak-2010} In a more complicated case of antiferromagnetic order, e.g. easy plane antiferromagnet with domains, angular dependence of torque will not be described by \eqref{eq:angularTorque}.\cite{Herak-2010} 
\subsubsection{Paramagnetic state}\label{sec:torquePM}
\begin{figure}[t]
	\centering
		\includegraphics[width=0.80\columnwidth]{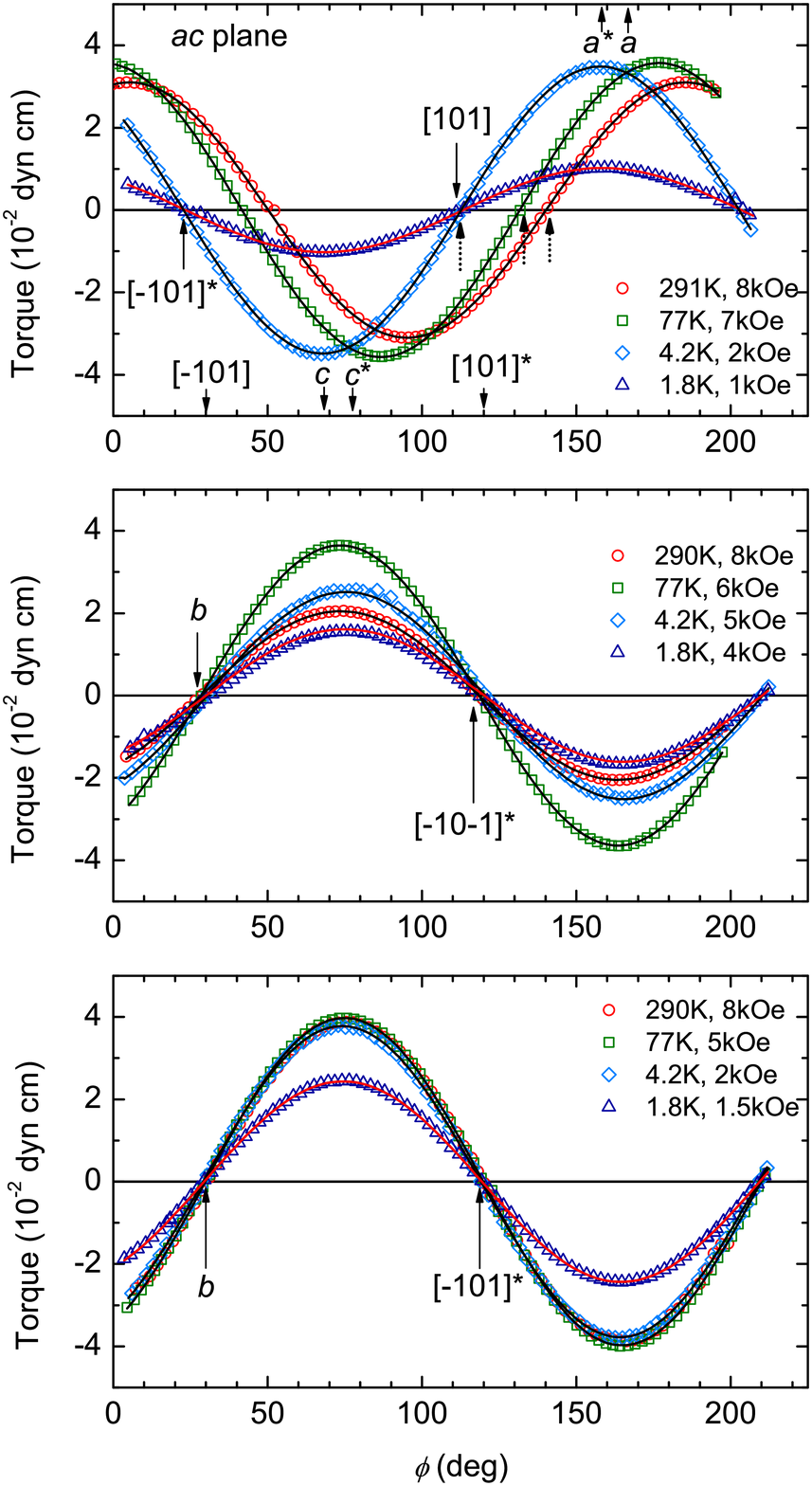}
	\caption{(Color online) Angular dependence of torque measured in three different crystal planes: $ac$ plane, $(b, [1 0 1]^*)$ plane and $(b, [\bar{1} 0 1]^*)$ plane. Solid lines represent fits to \eqref{eq:angularTorque}. Dotted arrows in top panel mark the position of $\phi_0$ at different temperatures.}
	\label{fig2}
\end{figure}

\begin{figure}[t]
	\centering
		\includegraphics[clip,width=0.80\columnwidth]{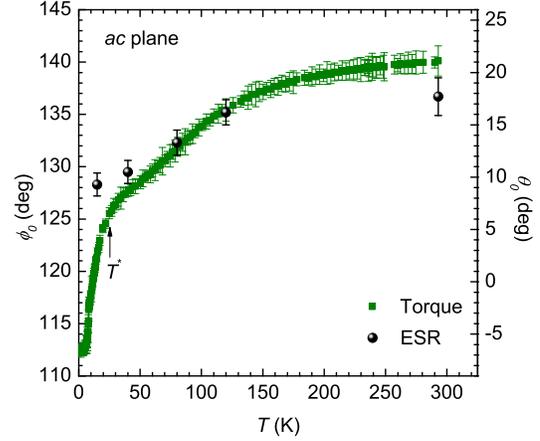}
		\caption{(Color online) Comparison of the temperature dependence of the shift of magnetic axes in the $ac$ plane measured by torque magnetometry and ESR. For meaning of $\phi_0$ see \eqref{eq:angularTorque} and of $\theta_0$ \eqref{eq:angularESR}. $T^*$ is a temperature at which a kink is observed in $\Delta \chi_{b,\: [1 0 1]^*}$ [see \fref{fig4}(b)].}
	\label{fig3}
\end{figure}

\begin{figure}[tb]
	\centering
		\includegraphics[clip,width=0.80\columnwidth]{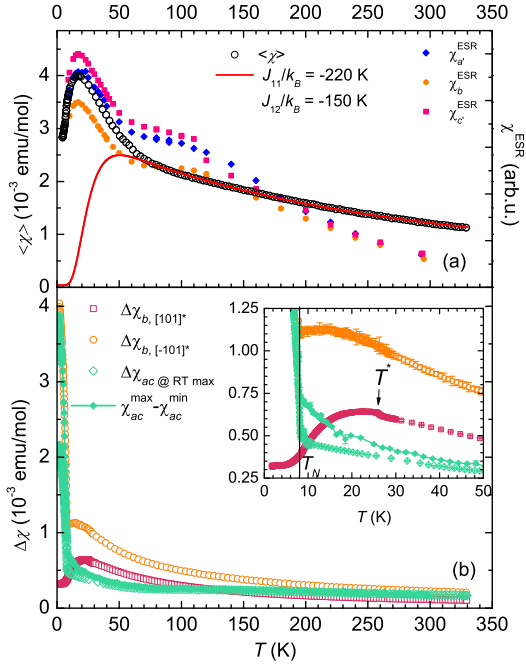}
	\caption{(Color online) Temperature dependence of (a) average magnetic susceptibility $<\chi>$ of \cso compared to measured ESR intensities along three mutually perpendicular directions and (b) susceptibility anisotropy measured in three crystal planes.}
	\label{fig4}
\end{figure}

\indent Angular dependence of torque measured at different temperatures in the three mentioned crystallographic planes is shown in \fref{fig2}. All measured curves can be well described by \eqref{eq:angularTorque}. Values of susceptibility anisotropy and phase $\phi_0$ obtained by fitting the measured data to \eqref{eq:angularTorque} are given in Table \ref{tab:phasetorque}. We see that in both planes containing axis $b$ the torque is zero along the $b$ axis. This means that the $b$ axis is one of the principal macroscopic magnetic axes in \cso in the entire temperature range. The other two principal magnetic axes are confined to the $ac$ plane. From the top panel of \fref{fig2}, it is clear that magnetic axes in the $ac$ plane rotate as the temperature changes, as was already reported previously.\cite{Zivkovic-2012} This rotation is represented by a temperature shift of the phase $\phi_0$ (see dotted arrows in \fref{fig2}). The temperature dependence of $\phi_0$ measured on the sample used in this work shown in \fref{fig3} is in complete agreement \footnote{Reader should not be confused by the different absolute values of $\phi_0$ given here and in Ref. \onlinecite{Zivkovic-2012}. The sample was differently oriented with respect to the goniometer angle $\phi=0$ in these two experiments.} with the result published previously.\cite{Zivkovic-2012} It should be mentioned, however, that in our previous work the position of the crystal axes in Fig. 7b) of Ref. \onlinecite{Zivkovic-2012} was incorrectly labeled. According to results we obtained here, the axis labeled $[\bar{1} 0 1]$ in Ref. \onlinecite{Zivkovic-2012} is in fact the $[1 0 1]$ axis, and the axis labeled $\tau = b\times [\bar{1} 0 1]$ Ref. \onlinecite{Zivkovic-2012} is in fact the $[\bar{1} 0 1]^*$ axis.\\
\begin{table}[tb]
\centering
		\begin{tabular*}{\linewidth}{@{\extracolsep{\fill}}c c c c }
		\hline \hline
			&	T~(K) & $\Delta \chi$~($10^{-3}$~emu/mol)	& $\phi_0$ ($\deg$)\\	
				\hline
			\multirow{4}{2cm}{$ac$ plane.}&	291 & 0.1793(1) & 140.2(2)\\
			&	77 & 0.2696(2) & 131.7(2)\\
			&	4.2 & 3.222(3) & 112.7(2)\\
			&	1.8 & 3.78(1) & 113.4(2)\\ \hline %\hline
			%	& & & \\
			%\hline \hline
			%
	%& T~(K) & $\Delta \chi$~($10^{-3}$~emu/mol)	& $\phi_0$ ($\deg$)\\	\hline
		\multirow{4}{2cm}{$(b, [1 0 1]^*)$ plane.}&	290 &  0.11184(1) & 28.5(2)\\
	&	77 & 0.3745(2) & 28.5(2)\\
	&	4.2 & 0.3723(7) & 30.3(2)\\
	&	1.8 & 0.372(1) & 30.1(2)\\ \hline %\hline	& & & \\
			%	\hline \hline
%&	T~(K) & $\Delta \chi$~($10^{-3}$~emu/mol)	& $\phi_0$ ($\deg$)\\	\hline
%
\multirow{4}{2cm}{$(b, [\bar{1} 0 1]^*)$ plane.}		&	290 & 0.1408(2) & 29.4(2)\\
	&	77 & 0.5883(3) & 29.9(2)\\
	&	4.2 & 3.496(2) & 29.1(2)\\
	&	1.8 & 4.005(5) & 29.4(2)\\ \hline \hline
		\end{tabular*}
\caption{Parameters obtained from fits of measured angular dependencies of torque shown in \fref{fig2} to \eqref{eq:angularTorque}.}
\label{tab:phasetorque}
\end{table}
\indent Temperature dependence of average magnetic susceptibility $<\chi>$ shown in \fref{fig4}(a) is compared to the susceptibility of a linear tetramer model\cite{Emori-1975} for parameter values $J_{11}/k_B= - 220$~K, $J_{12}/k_B = -150$~K, $\chi_0 = 4\cdot 10^{-5}$~emu/mol and by setting $<g>=2.19$ (value taken from our ESR measurements, section \ref{sec:gtensor}), very similar to the values reported previously.\cite{Zivkovic-2012} The observed disagreement between theory and experiment in principle might be a result of the temperature change of $g$ factor\cite{Zivkovic-2012} which is not taken into account in the theory. Our detailed ESR results presented in Section \ref{sec:esr} will help us clarify this. \\
\indent Temperature dependence of magnetic susceptibility anisotropies in the $ac$ plane, $(b, \:[\bar{1} 0 1]^*)$ plane, and $(b, \: [1 0 1]^*)$ plane, $\Delta \chi_{ac}$, $\Delta \chi_{b,\: [\bar{1} 0 1]^*}$, and $\Delta \chi_{b,\: [1 0 1]^*}$, respectively, is shown in \fref{fig4}(b). Because the magnetic axes rotate in the $ac$ plane, we performed two types of measurement in that plane. The data for $\Delta \chi_{ac\: @RT\:max}$, $\Delta \chi_{b,\: [\bar{1} 0 1]^*}$, and $\Delta \chi_{b,\: [1 0 1]^*}$ were obtained with magnetic field applied at the angle of the maximum of sine curve at room temperature (RT) (see \fref{fig2}). Since phase $\phi_0$ in the $(b,\: [\bar{1} 0 1]^*)$ and $(b,\: [1 0 1]^*)$ planes does not change with temperature, this type of measurement gives the susceptibility anisotropy in the entire temperature range. That is not so for the $ac$ plane, therefore, we also performed measurements which simultaneously measure temperature dependence of $\phi_0$ and the torque amplitude. This procedure gives anisotropy $\chi_{ac}^{max} - \chi_{ac}^{min}$. Only at lowest temperatures is this anisotropy equal to $\chi_{[101]} - \chi_{[\bar{1} 0 1]^*}$ (see top panel of \fref{fig2}). The difference between $\Delta \chi_{ac\: @RT\:max}$ and $\chi_{ac}^{max} - \chi_{ac}^{min}$ clearly increases at temperatures at which the shift of $\phi_0$ becomes significant, as can be seen in the inset of \fref{fig4}(b). $\Delta \chi_{b,\: [\bar{1} 0 1]^*}$ has a maximum at $\approx 13$~K and only slightly decreases as the temperature decreases toward $T_N$. Phase transition to the ordered state at $T_N$ is observed as a kink below which $\Delta \chi_{b,\: [\bar{1} 0 1]^*}$ increases as the temperature decreases. $\Delta \chi_{b,\: [1 0 1]^*}$ has a maximum at $\approx 23$~K and then significantly decreases as $T$ decreases to $T_N$. $\Delta \chi_{ac}$ increases less rapidly as temperature decreases than the other two anisotropies. There is an indication of a wide maximum around $\approx 125$~K, but below $\approx 100$~K $\Delta \chi_{ac}$ starts to increase as the temperature decreases down to $T_N$, and no maximum is observed at low temperatures above $T_N$. These different temperature dependencies of susceptibility anisotropies measured in different planes are inconsistent with the model of isotropic Heisenberg Hamiltonian with temperature-independent $\bm{g}$ tensor. We explained at the beginning of this section how temperature dependence of the $\bm{g}$ tensor has direct influence on the susceptibility anisotropy. To determine if this is the (only) reason for the observed behavior of susceptibility anisotropy, we performed detailed ESR measurements which will be presented in Sec. \ref{sec:esr}.\\
\indent One more feature not observed previously is also reported here in the inset of \fref{fig4}(b). At $T^*\approx 26$~K, $\Delta \chi_{b,\: [1 0 1]^*}$ has a visible kink which is not observable in the other two planes. We performed these measurements several times to make certain that this is not an experimental artifact. This is also roughly the temperature below which a steep change of $\phi_0$ is observed in the $ac$ plane (see \fref{fig3}). While this transition does not represent a phase transition to long-range magnetically ordered state, its nature is not clear otherwise. Structural phase transition of some kind could be one possible explanation. At this moment, however, we can only speculate about its origin, so we leave the clarification of the meaning of $T^*$ for future experimental studies.
\subsubsection{Magnetically ordered state}\label{sec:torqueAFM}
\begin{figure}[tb]
	\centering
		\includegraphics[clip,width=0.80\columnwidth]{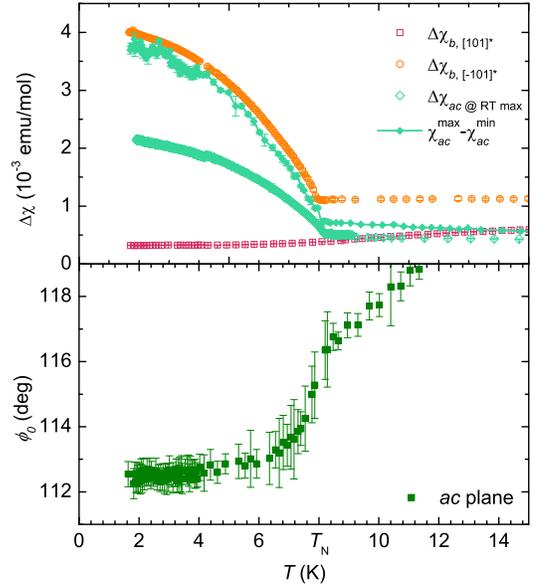}
		\caption{(Color online) Top: temperature dependence of susceptibility anisotropy in the magnetically ordered state. Bottom: Temperature dependence of phase $\phi_0$ in the $ac$ plane [see \eqref{eq:angularTorque}] in the magnetically ordered state.}
	\label{fig5}
\end{figure}

\indent Temperature dependence of the susceptibility anisotropy in the vicinity and below $T_N$ is shown in the top panel of \fref{fig5}. Phase transition to magnetically ordered state is clearly visible as a kink at $T_N\approx 8.2$~K in the $ac$ and $(b,\: [\bar{1} 0 1]^*)$ planes, while it is hardly observable in the $(b,\: [1 0 1]^*)$ plane. In the $ac$ and the $(b,\: [\bar{1} 0 1]^*)$ planes susceptibility anisotropy increases as the temperature decreases, while in the $(b,\: [1 0 1]^*)$ plane it slightly decreases with decreasing temperature. $\Delta \chi_{ac\:@RT\:max}$ and $\chi_{ac}^{max}- \chi_{ac}^{min}$ have the same meaning as already discussed. This behavior of susceptibility anisotropy can be mapped to the temperature dependence of susceptibility typical for the Ne\'{e}l collinear antiferromagnet\cite{AM} with easy axis along the $[\bar{1} 0 1]^*$ direction, where $\chi_{\|}=\chi_{[\bar{1} 0 0]^*}$, $\chi_{IM}=\chi_{[1 0 1]}$ and $\chi_{\perp}=\chi_b$. However, small canting of spins is not excluded, as we discuss below.\\
\indent The bottom panel of \fref{fig5} shows temperature dependence of the phase $\phi_0$ in the $ac$ plane in the same temperature range. The phase, i.e., the position of magnetic axes also displays a kink at $T_N$ and continues to change with temperature even below $T_N$. Total rotation of magnetic axes below $T_N$ in the $ac$ plane amounts to 3$\deg$. The rotation stops  only below $\approx 4$~K and it is only below that temperature that the $[\bar{1} 0 1]^*$ and $[101]$ axes become magnetic axes. At those temperatures $\chi_{ac}^{max}- \chi_{ac}^{min} = \chi_{[101]} - \chi_{[\bar{1} 0 1]^*}$.\\
\indent Observed behavior of susceptibility anisotropy is in agreement with collinear antiferromagnetic order with easy axis along $\approx <\bar{1} 0 1>^*$. The spins are either collinear and directed along $\approx <\bar{1} 0 1>^*$ axes, or the dominant and large projection of spins is along that direction. Small canting of spins on tetramer is not excluded, but it cannot be observed by our bulk measurements. However, all our angular dependencies measured in the ordered state (see \fref{fig2}) are described by \eqref{eq:angularTorque}, which means that there is no bulk weak ferromagnetism present in the ordered state.\cite{Herak-2010,Herak-2013} Thus, the small canting of spins, if it exists, must be such to produce no net macroscopic magnetic moment. This is corroborated by symmetry: the same inversion center exists between spins $\bm{S}_2$ and $\bm{S}_3$ and spins $\bm{S}_1$ and $\bm{S}_4$ and canting is possible only between spins $\bm{S}_1$ and $\bm{S}_2$ and spins $\bm{S}_3$ and $\bm{S}_4$ (see \fref{fig1}). Thus, assuming AFM interactions between spins in Hamiltonian \eref{eq:Hamiltonian}, spin arrangement must be such to produce no net magnetic moment on tetramer.
%
%
%%%%%%%%%%%%%%%%%%%%%%%%%%%%%%%%%%%%%%%%%%%%%%%%%%%%%%%%%%%%%%%%%%%%%%%%%%%%%%%%%%%%%%%%%%%
%    
%                                ESR
%
%%%%%%%%%%%%%%%%%%%%%%%%%%%%%%%%%%%%%%%%%%%%%%%%%%%%%%%%%%%%%%%%%%%%%%%%%%%%%%%%%%%%%%%%%%%
%
\subsection{Electron spin resonance}\label{sec:esr}
\begin{figure}[b]
	\centering
		\includegraphics[width=0.80\columnwidth]{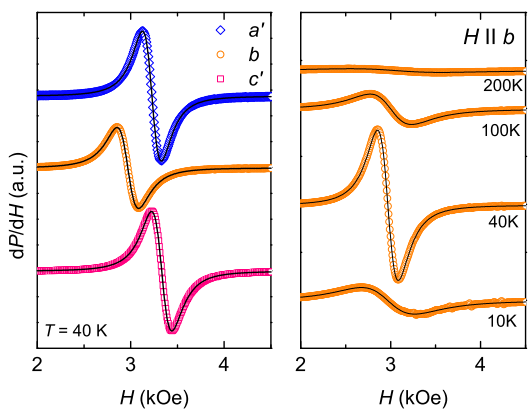}
	\caption{(Color online) ESR spectra in \cso measured at $T=40$~K for magnetic field parallel to mutually perpendicular $a'$, $b$, and $c'$ axes (left panel) and at different temperatures for $\mathbf{H} \parallel b$ (right panel). Solid lines represent fits to single Lorentzian line shape.}
	\label{fig6}
\end{figure}

\indent We measured the temperature dependence of X-band ESR spectra in the $ac$ and $(b,\:  [1 0 1]^*)$ planes in the temperature range from 9~K to room temperature. In all measurements field sweep was in the range from 500 to 7500~Oe. At all measured temperatures and for all field directions, we observed a single Lorentzian line. Below 9~K, the ESR signal disappears as the system enters a magnetically ordered state. Spectra at 40~K for $H$ parallel to three mutually perpendicular directions $a'$, $b$, and $c'$ are shown in the left panel of \fref{fig6}. Axes $a'$ and $c'$ in the $ac$ plane were erroneously assigned as axes $a^*$ and $c$ during ESR measurements, but our structural measurements later revealed that these are in fact $6.7\deg$ away from the $a^*$ and $c$ axes in the same plane. The temperature evolution of spectra for $H \parallel b$ is shown in the right panel of the same figure. All the spectra were fitted to a single Lorentzian line shape which is represented by a solid line in \fref{fig6}. In \fref{fig4}(a), we show ESR intensities obtained by doubly integrating the obtained spectra. The behavior of ESR intensities is similar to the measured susceptibility, especially at low temperatures where both $<\chi>$ and $\chi^{ESR}$ display maximum at approximately same temperature.
\subsubsection{\texorpdfstring{$\bm{g}$ tensor}{g tensor}}\label{sec:gtensor}
\begin{figure}[t]
	\centering
		\includegraphics[clip,width=0.80\columnwidth]{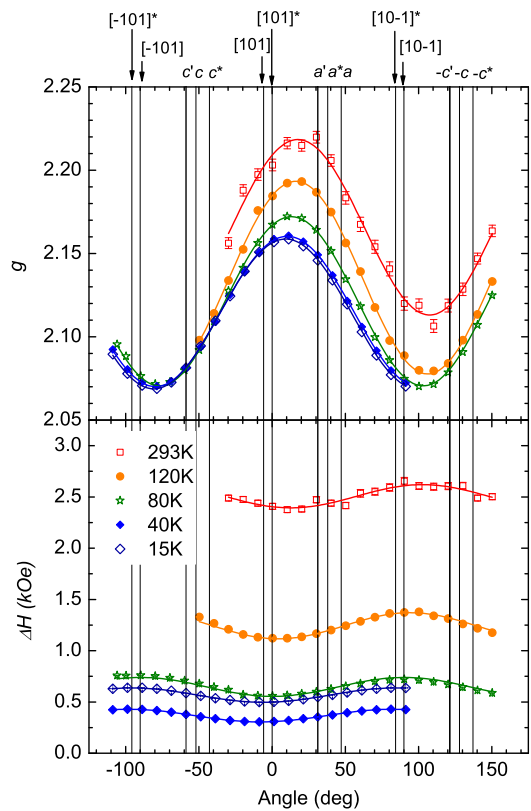}
		\caption{(Color online) Angular dependence of $g$ factor (top panel) and ESR linewidth (bottom panel) in the $ac$ plane measured at several different temperatures.  Solid lines represent fits to \eqref{eq:angularESR} (top)  and \eqref{eq:lwangular} (bottom).}
	\label{fig7}
\end{figure}

\begin{figure}[t]
	\centering
		\includegraphics[clip,width=0.80\columnwidth]{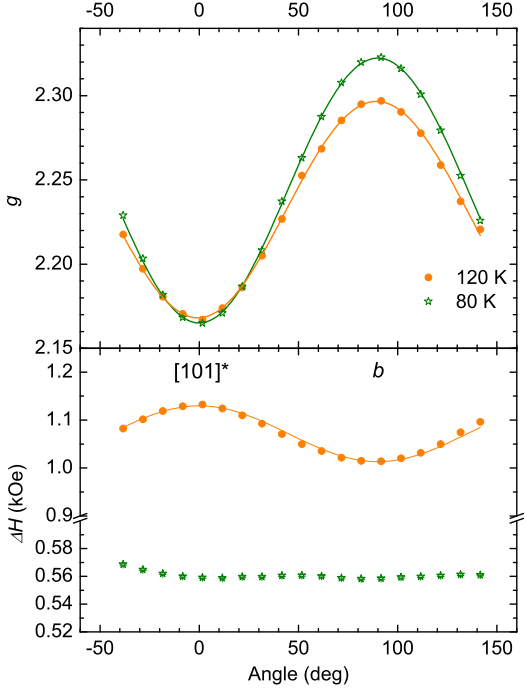}
		\caption{(Color online) Angular dependence of $g$ factor (top panel) and ESR linewidth (bottom panel) in the $(b,\: [1 0 1]^*)$ plane measured at 120~K and 80~K. Solid lines represent fit to \eqref{eq:angularESR} (top)  and \eqref{eq:lwangular} (bottom).}
	\label{fig8}
\end{figure}

\indent Angular dependence of $\bm{g}$ tensor in the $ac$ plane measured at temperatures of 293, 120, 80, 40, and 15~K is shown in \fref{fig7}. Because of the sample morphology ,we chose the direction of axis $[101]^*$ as the goniometer angle 0$\deg$, and axis $[10\bar{1}]$ as 90$\deg$. Note that these directions correspond to different goniometer angle values in torque measurements. It is immediately obvious that the $\bm{g}$ tensor rotates and changes magnitude as the temperature decreases. The total $\bm{g}$ tensor obtained at different temperatures is calculated in the Appendix \ref{sec:app1} and the results are given in Table \ref{tab:gcomponents}. Since we want to compare the rotation of the $\bm{g}$ tensor to the rotation of macroscopic magnetic axes obtained form torque, we fit the measured values to the following expression 
\begin{equation}\label{eq:angularESR}
	g = \sqrt{g_{max}^2 \cos^2(\theta - \theta_0)+ g_{min}^2 \sin^2(\theta - \theta_0)}
\end{equation}
where $g_{max}$ and $g_{min}$ are maximal and minimal values of $\bm{g}$ tensor in the plane of measurement and $\theta_0$ is the angle at which $g_{max}$ is measured. $\theta$ is the goniometer angle. Results are summarized in Table \ref{tab:phaseESR}. The errors are obtained from the fitting procedure, but we must also take into account the goniometer uncertainty of $\approx 1\deg$. Total change in $\theta_0$ between 120 and 15~K is $\approx 7\deg$. This is in agreement with previously published results where angular dependence in the $ac$ plane was measured at 100 and 15~K.\cite{Zivkovic-2012} In \fref{fig3}, we plot the results for $\theta_0$ given in Table \ref{tab:phaseESR} against the phase change $\phi_0$ measured by torque in the same plane. As we can see from the figure, the rotation of magnetic axes observed in torque agrees reasonably well to the rotation of $\bm{g}$ tensor at temperatures $T\gtrsim 50$~K. Below that temperature, the rotation of the $\bm{g}$ tensor becomes insignificant when compared to the large rotation of magnetic axes observed in torque.\\
\begin{table}[tb]
\centering
		\begin{tabular*}{\linewidth}{@{\extracolsep{\fill}}c c l l l}
		\hline \hline
			&	T~(K) & $g_{max}$	& $g_{min}$ &$\theta_0$($\deg$)\\	
				\hline
			\multirow{5}{*}{$ac$ plane.}&	293 & 2.218(2) & 2.113(2) & 17.7(8) \\
			&	120 & 2.1934(6) & 2.0777(6) & 16.2(2) \\
			&	80 & 2.1721(4) & 2.0705(3) & 13.3(2) \\
			&	40 & 2.1603(2) & 2.0701(1) & 10.5(1) \\ 
			&	15 & 2.1587(2) & 2.0687(1) & 9.3(1) \\ 
			\hline \hline	& & & &\\ 
				\hline \hline
	& T~(K) & $g_{b}$ & $g_{[1 0 1]^*}$	& $\theta_0$ ($\deg$)\\	\hline
		\multirow{2}{2cm}{Plane with axes $b$ and $ [1 0 1]^*$}&	120 &  2.2967(6) & 2.1681(6) & 0\\
	&	80 & 2.3224(4) & 2.1651(4) & 0\\ 
			 \hline \hline
		\end{tabular*}
\caption{Parameters obtained from fits of measured angular dependences of $g$ factor to \eqref{eq:angularESR}.}
\label{tab:phaseESR}
\end{table}
\indent Angular dependencies of $g$ factor in $(b,\:  [1 0 1]^*)$ and $(b,\: [\bar{1} 0 1]^*)$ plane at temperatures 100 and 15~K were reported previously.\cite{Zivkovic-2012} No phase shift was observed in these planes. In \fref{fig8}, we show angular dependence of $g$ factor and linewidth in the $(b,\: [1 0 1]^*)$ plane measured at 120 and 80~K, and the results of fit to \eqref{eq:angularESR} with $g_{max}=g_b$ and $g_{min}=g_{[1 0 1]^*}$ are given in Table \ref{tab:phaseESR}. Within the experimental error, we observe no rotation of the $\bm{g}$ tensor, as reported previously. This is also in agreement with the torque results which show no rotation of magnetic axes in paramagnetic state in this plane.\\
\begin{figure}[t]
	\centering
		\includegraphics[clip,width=0.80\columnwidth]{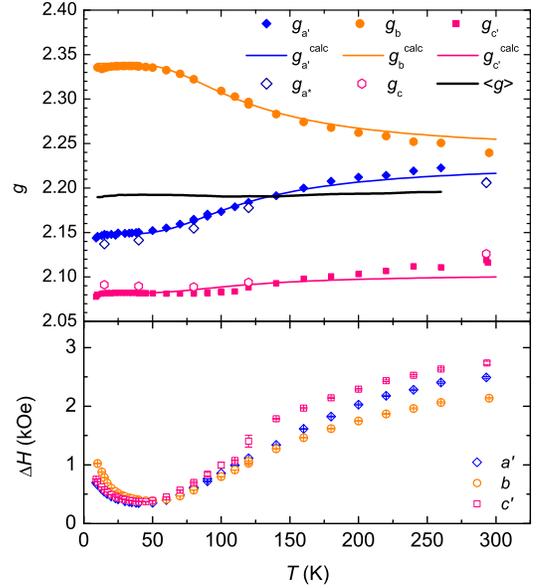}
		\caption{(Color online) Temperature dependence of $\bm{g}$ tensor and $<g>$ (top panel) and ESR linewidth (bottom panel) measured for $H \parallel a',\;b, \; c'$. Values of $g_{a*}$ and $g_c$ taken from angular dependences (\fref{fig7}) are also shown. Solid lines represent results of calculations for $J_{11}/k_B=290~K$ presented in Appendix \ref{sec:app1} (see text).}
	\label{fig9}
\end{figure}
\indent In \fref{fig9} we show temperature dependence of the $g$ factor measured for $H$ parallel to three mutually perpendicular axes $a'$, $b$, and $c'$. We also show the average $g$ factor, $<g> = [1/3\: (g_{a'}^2+g_b^2+g_{c'}^2)]^{1/2}=2.19$, and values of $g_{a^*}$ and $g_c$ taken from angular dependences shown in \fref{fig7}. As can be seen in \fref{fig9}, average $g$ factor is temperature independent. This finding, together with results shown in Figs. \ref{fig3} and \ref{fig4}, suggests that apart from the $g$-factor anisotropy other contributions to magnetic anisotropy, such as symmetric and antisymmetric anisotropic exchange interaction (DMI), have significant influence on the ground state of \csos. Thus, we next present the analysis of the ESR linewidth which is very sensitive to the presence of anisotropic exchange interactions.
\subsubsection{ESR linewidth}\label{sec:esrlw}
\begin{figure}[t]
	\centering
		\includegraphics[width=0.80\columnwidth]{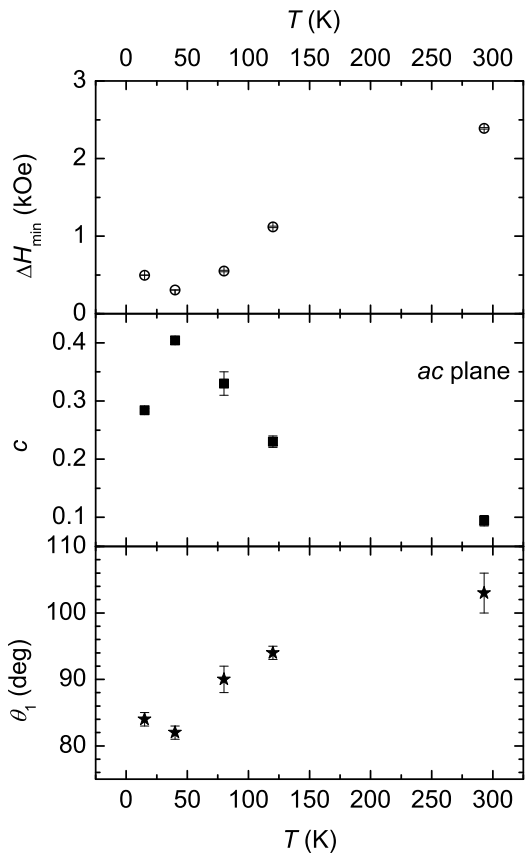}
	\caption{Temperature dependence of the parameters obtained from the fit of angular dependence of linewidth in the $ac$ plane to \eqref{eq:lwangular}.}
	\label{fig10}
\end{figure}

\indent Angular dependence of the ESR linewidth $\Delta H$ in the $ac$ and  $(b,\: [1 0 1]^*)$ planes is shown in Figs. \ref{fig7} and \ref{fig8}, respectively. ESR linewidth is anisotropic and its anisotropy significantly changes with temperature. In order to obtain some information about the temperature dependence of the anisotropy of the linewidth, we fitted the measured angular dependencies to the expression used for the dimer system\cite{Camara-2010}
\begin{equation}\label{eq:lwangular}
	\Delta H =\Delta H_{min}[1+c\: \cos^2(\theta - \theta_1)].
\end{equation}
Factor $c$ is connected to the anisotropic interactions. $\theta_1$ represents the angle at which the linewidth has maximal value. This approach has been successfully used to describe angular dependence of linewidth in quasi-two-dimensional (quasi-2D) spin system where the maximal linewidth was observed at the angle of minimal $g$ and vice versa.\cite{Soos-1977,Zhou-1992} Since our system is different than those for which \eqref{eq:lwangular} was obtained, we use \eqref{eq:lwangular} empirically in order to describe the change of phase $\theta_1$ with temperature. The results of fit are shown in Table \ref{tab:phaseDeltaH} and we plot the temperature dependence of the parameters obtained from the fit in \fref{fig10}. Temperature dependence of $\Delta H_{min}$ roughly corresponds to the measured temperature dependence of the linewidth shown in \fref{fig10} with minimum observed around 40~K. Coefficient $c$ increases as the temperature decreases and then below $\approx 40$~K it starts to decrease with decreasing temperature. The most interesting result is that in the $ac$ plane not only do the minima and maxima of $\Delta H$ not correspond to the minima and maxima of the $g$ factor, but the phase $\theta_1$ in \eref{eq:lwangular} changes with temperature more rapidly than the phase $\theta_0$ of the $\bm{g}$ tensor (Table \ref{tab:phaseESR}). Total change of $\theta_1$ for $\Delta H$ amounts to $\approx 20$~$\deg$ between room temperature and 15~K, while for the $\bm{g}$ tensor it only amount to $\approx 9$~$\deg$ in the same temperature range. In the $(b,\: [1 0 1]^*)$ plane at 120~K the angle of the maximal linewidth corresponds to the angle of minimal $g$ factor and vice versa. At $T=80$~K the angular dependence of the linewidth is very weak and does not seem to be described by \eqref{eq:lwangular}. The temperature dependence of $\theta_1$ strongly suggests that there are at least two different contributions affecting the ESR linewidth.\\
\begin{table}[tb]
\centering
		\begin{tabular*}{\linewidth}{@{\extracolsep{\fill}}c c c c c}
		\hline \hline
			&	T~(K) & $\Delta H_{min}$ (kOe)	& $c$ &$\theta_1$($\deg$)\\	
				\hline
			\multirow{5}{*}{$ac$ plane.}&	293 & 2.39(1) & 0.094(9) & 103(3) \\
			&	120 & 1.12(1) & 0.23(1) & 94(1) \\
			&	80 & 0.55(1) & 0.33(2)& 90(2) \\
			&	40 & 0.305(1) & 0.404(3) & 82(1) \\ 
			&	15 & 0.637(1) & 0.284(3) & 84(1) \\ 
				\hline %\hline	& & & &\\ 
				%\hline \hline
				 	\multirow{2}{2cm}{Plane with axes $b$ and $[1 0 1]^*$} & 120 & 1.01(1) & 0.115(4) & 0(1)\\
				&80  & -- & -- & -- \\
			\hline \hline	
				\end{tabular*}
\caption{Parameters obtained from fits of measured angular dependences of ESR linewidth $\Delta H$  to \eqref{eq:lwangular}.}
\label{tab:phaseDeltaH}
\end{table}
\indent Temperature dependence of the ESR linewidth $\Delta H$ measured for $H\| a'$, $b$, and $c'$ is shown in \fref{fig9}. $\Delta H$ decreases as the temperature decreases down to $\approx 45$~K where it has a minimum and then increases as the temperature decreases. At high temperatures $\Delta H_b< \Delta H_{a'} < \Delta H_{c'}$ while below 45~K $\Delta H_c' \approx \Delta H_{a'} < \Delta H_b$. This also suggests several different contributions to the ESR linewidth. Next, we attempt to distinguish different contributions.\\
\indent In low-dimensional compounds, ESR linewidth is broader compared to three-dimensional systems.\cite{Bencini-1990} For the  latter, an expression for the observed linewidth $\Gamma \approx M_2/\Gamma_{exch}$ is valid, where $\Gamma$ is observed peak-to-peak half-width given in Gauss units, $M_2$ is the second moment, and $\Gamma_{exch}$ is the strength of exchange interaction.\cite{Goldman-1970} For one-dimensional compounds, the
width of expected exchange-narrowed line could be calculated as\cite{Bencini-1990}
\begin{equation}
\Gamma \approx M_2^{2/3}/\Gamma_{exch}^{1/3}. \label{eq_linewidth}
\end{equation}
The total second moment $M_2$ is the sum of different broadening contributions\cite{Bencini-1990}
\begin{equation}
M_2 = M_2^{dip} + M_2^{\Delta g} +  M_2^{hyp} + M_2^{an.exch}.
\label{eq_M2sum}
\end{equation}
$M_2^{dip}$ is the second moment arising from dipole-dipole interaction and other terms $M_2^{\Delta g}$, $M_2^{hyp}$, and $M_2^{an.exch}$  originate from other anisotropic interactions: anisotropic Zeeman interaction, hyperfine coupling and anisotropic
exchange interaction, respectively. In order to determine the dominant interaction responsible for the broad ESR line of the SeCuO$_3$ system, we have calculated and estimated individual contributions of all terms in \eqref{eq_M2sum}.\\
\indent The second moment arising from dipole-dipole interaction $M_2^{dip}$, in approximation for a powder sample with cubic symmetry, could be obtained as\cite{VanVleck-1948}
\begin{equation}
M_2^{dip} = \frac{3}{5} g^2 \mu_B^2 S(S+1)\sum_{j}r_{ij}^{-6},
\label{eq_M2}
\end{equation}
where, of course, the term $i=j$ is excluded from the summation, $g=2.19$ is the average $g$ value taken from the experiment, and other symbols have their usual meaning. Using \eqref{eq_M2}, it has been found that the dipolar contribution from all copper ions located in the sphere of radius 12 {\AA} is $M_2^{dip} \approx 6.8\cdot 10^5$~G$^2$. The sum in \eqref{eq_M2} converged very fast, therefore the taken radius was large enough for calculation.\\
\indent The anisotropic Zeeman contribution $M_2^{\Delta g}$ is a result of the existence of four magnetically non equivalent copper ions in SeCuO$_3$ (two different spatial orientation of Cu1 and Cu2 ions). This contribution to the ESR linewidth is generally estimated as\cite{Bencini-1990}
\begin{equation}
\Gamma^{\Delta g} \approx (\mu_B /8gJ)[(g_1-g_2)H_{res}]^2.
\label{eq_anisZeeman}
\end{equation}
If we take the maximum value of $|g_1-g_2|=0.4$ from the $g$ factor calculation given in the Appendix and interaction between inequivalent Cu1 and Cu2 $J/k_B=|J_{12}/k_B|=150$~K from the linear tetramer model of susceptibility, the value $\Gamma^{\Delta g}=0.04$~G is obtained. Therefore, the anisotropic Zeeman term could be neglected in the total the X-band linewidth while its contribution is significant in the Q band and at higher frequencies.\\
\indent Another relevant contribution originates from hyperfine coupling of electron with nuclear spin $I=3/2$.
\begin{align}
\nonumber
M_2^{hyp} &= [I(I+1)/3] K^2(\theta)/g^2(\theta) + \\
&+[I(I+1)/6][A_{\|}^2 A_{\bot}^2 /g^2(\theta)+A_{\|}^2+E^2(\theta)].
 \label{eq_hyperfine}
\end{align}
$A_{\|}$ and $A_{\bot}$ are parallel and perpendicular principal components of axial hyperfine $\mathbf{A}$ tensor and other terms are defined in Ref.\cite{Bencini-1990}. If we take the usual experimental value for Cu ion, $A_{\|}=200$~G and $A_{\bot}=0$, the hyperfine contribution to the second moment could be maximally estimated as $M_2^{hyp} \approx 7.5\cdot 10^4$~G$^2$. Compared to the value $M_2^{dip} \approx 6.8\cdot 10^5$~G$^2$, one can see that hyperfine contribution is an order of magnitude smaller
compared to the dipolar contribution to the second moment.\\
\indent Therefore, the second moment originating from dipolar, hyperfine and neglectingly small anisotropic Zeeman interaction amounts to $M_2 = M_2^{dip} + M_2^{\Delta g} + M_2^{hyp}=7.55\cdot 10^5$~G$^2$. The strength of exchange interaction $J/k_B=150$~K expressed in Gauss units is $\Gamma_{exch}\approx 10^6$ G. Using Eq.~\ref{eq_linewidth}, the width of expected exchange narrowed line $\Gamma = 66$~G is obtained and the full-width at half-height\cite{Weil} $\Delta H = (2\:\sqrt{2ln2})\:\Gamma = 155 $~G. This value is significantly smaller compared to the smallest experimental linewidth observed $\Delta H_{min}=377$~G (Fig.~10). The obtained difference in linewidth (with minimum value $377-155\sim200$~G) must be produced by anisotropic exchange interaction. Therefore, from the previous linewidth estimation, one can conclude that symmetric and/or antisymmetric anisotropic exchange interaction is present in SeCuO$_3$ system.
%
%
%
%%%%%%%%%%%%%%%%%%%%%%%%%%%%%%%%%%%%%%%%%%%%%%%%%%%%%%%%%%%%%%%%%%%%%%%%%%%%%%%%%%%%%%%%%%
%
%         DISCUSSION AND CONCLUSION
%
%%%%%%%%%%%%%%%%%%%%%%%%%%%%%%%%%%%%%%%%%%%%%%%%%%%%%%%%%%%%%%%%%%%%%%%%%%%%%%%%%%%%%%%%%%
%
%
\section{Discussion}\label{sec:disc}
\indent Before going to the main point of this work, we discuss the possible origin of the observed temperature dependence of the $\bm{g}$ tensor. In principle, it can be a consequence of the structural changes of the ligand surrounding of the magnetic ion or a dynamic Jahn-Teller distortion.\cite{Kiczka-2004,Hoffmann-2010} However, no significant temperature change of the crystal structure of \cso was observed in the temperature range 80-300~K.\cite{Zivkovic-2012} The possibility of a dynamic Jahn-Teller distortion and freezing of the system in some configuration with lowest energy at low temperatures can not be excluded. Dynamic Jahn-Teller effect is a reason behind temperature change of the $\bm{g}$ tensor observed in CuSb$_2$O$_6$ where a crossover takes place from high-temperature tetragonal to low-temperature monoclinic crystal structure in a wide temperature range finally resulting in the freezing of the system below $\approx 200$~K.\cite{Heinrich-2003} When CuSb$_2$O$_6$ crystal structure freezes, the crystal usually ends up having several twins due to several possibilities for the orientation of CuO$_6$ octahedra. This is reflected in ESR spectra as a presence of several lines at different resonant fields. In contrast, only one single-Lorentzian ESR line is observed in \cso at all temperatures. While dynamic Jahn-Teller distortion averages the anisotropic components of the $\bm{g}$ tensor at high temperatures, the directions of the principal axes of the $\bm{g}$ tensor do not change with temperature.\cite{Hoffmann-2010} Furthermore, the symmetry of the CuO$_6$ octahedra in \cso is already low and one does not expect any further symmetry lowering. All this supports the idea put forward in Ref. \onlinecite{Zivkovic-2012} that temperature change of $\bm{g}$ tensor is not a consequence of the structural changes of crystal symmetry, however, high-resolution investigations of the temperature dependence of the crystal structure are needed to clarify this.\\
\indent In Ref.~\onlinecite{Zivkovic-2012}, it was suggested that the observed temperature dependence of the $\bm{g}$ tensor is a result of strong antiferromagnetic coupling of $J_{11}$ between central spins $\bm{S}_2$ and $\bm{S}_3$(see \fref{fig1}) which reduces their intensity in the total ESR signal. These spins are crystallographically inequivalent to $\bm{S}_1$ and $\bm{S}_4$ and thus have different principal $\bm{g}$ tensors. In this model of strongly coupled dimer, it was assumed that $J_{12}=0$ and $J_{11}=290$~K. The observed temperature dependence of two $\bm{g}$-tensor components presented in previous work was successfully simulated using the mentioned assumption. In the Appendix we further explore this proposal by calculating temperature and angular dependencies of the $\bm{g}$ tensor under the same assumptions using the measured $\bm{g}$ tensor and setting $J_{11}/k_B=290$~K, as in Ref. \onlinecite{Zivkovic-2012}. We see from \fref{fig9} that temperature dependence of measured $\bm{g}$-tensor components can be rather well described by this procedure. In the Appendix, we show that all our measured data can be successfully described in that manner. However, in this way we can neither obtain the principal values of $\bm{g}$ tensors nor their principal axes. We made an attempt to obtain these by defining the principal axes of each magnetically inequivalent Cu in the manner usually used for nondistorted CuO$_6$ octahedra, but in that case we found it impossible to describe our data using physically allowed values for $\bm{g}$-tensor components for Cu$^{2+}$ spin $S=1/2$. This result partially signifies that the distortion of CuO$_6$ octahedra which we ignored in our calculation is important, but also that the results obtained from simulation, although they give very good agreement with experiment, should not be taken without some reserve. 
 The main reason is that in calculations we assume that $J_{12}\ll J_{11}$, i.e $J_{12}/k_B$ should be of an order of 10~K or less. However, the Cu1-O-Cu2 angle in \cso is 108.5$\deg$ which is larger than in e.g. GeCuO$_3$ where $J/k_B\approx 100$~K. We thus expect that $J_{11}$ and $J_{12}$ are of the same order of magnitude. Further theoretical and experimental investigations of exchange coupling in \cso are necessary to clarify this.\\
\indent We now come to the main point of this work. We mentioned that the observed disagreement of measured $<\chi>$ with the magnetic susceptibility of the spin tetramer system described by the Hamiltonian \eref{eq:Hamiltonian} (\fref{fig4}) could have its origin in the temperature dependence of the $\bm{g}$ tensor. However, our ESR results show that $<g>$ is temperature -- independent and the reason for disagreement thus must lie elsewhere. A most trivial possibility is that the tetramer model does not describe the magnetic lattice of \csos. Although this would be surprising, it is not impossible. Theoretical analysis of superexchange paths and orbital ordering in \cso should help clarify this. This also leads us back to the possibility of some hard - to - detect structural change. Values of dominant interaction energies also remain to be determined. Further experimental and theoretical efforts are needed to resolve these issues.\\
\indent In a system with purely isotropic interactions between spins $S=1/2$ such as our proposed Hamiltonian \eref{eq:Hamiltonian} for \csos, magnetic anisotropy is a consequence of the $\bm{g}$ tensor anisotropy. As already mentioned, susceptibility anisotropy $\Delta\chi_{xy} \propto g_x^2-g_y^2$ and the rotation of the $\bm{g}$ tensor is reflected in the rotation of the susceptibility tensor which can be easily observed from torque measurements as the temperature change of the phase $\phi_0$ [see \eqref{eq:angularTorque}]. From \fref{fig3} we see that observed rotation of magnetic axes in torque in the $ac$ plane agrees reasonably well with the rotation of the $\bm{g}$ tensor only for $T \gtrsim 50$~K, which allows us to conclude that the rotation of $\bm{g}$ tensor is the probable cause of the rotation of macroscopic magnetic axes at these temperatures. Below $T\approx 50$~K the $\bm{g}$ tensor is temperature -- independent (\fref{fig9}). This means that large rotation of magnetic axes observed in torque below that temperature must have some other origin, but also that different temperature dependences of $\Delta \chi$ in different planes come from another source of anisotropy and not from the $\bm{g}$ tensor. The only other source of anisotropy for Cu$^{2+}$ spin $S=1/2$ system is the anisotropy of exchange interactions, either symmetric or antisymmetric or both. \\ 
\indent To see how anisotropic interactions influence susceptibility anisotropy $\Delta \chi$ we focus on the temperature range where the $\bm{g}$ tensor is temperature independent. For a spin $S=1/2$ system with isotropic interactions we have
\begin{equation}\label{eq:deltachiiso}
	\Delta \chi_{xy} =\chi_x(T) - \chi_y(T)\propto (g_x^2 - g_y^2)\: f(T),
\end{equation}
where $f(T)$ describes temperature dependence of susceptibility and its anisotropy. $f(T)$ is direction -- independent in the sense that susceptibility measured in any direction has the same $f(T)$ and the susceptibility magnitude is defined by magnitude of $g$ in that direction. $f(T)$ is defined by the Hamiltonian of the system. As we previously discussed and as can be seen from inset in \fref{fig4}(b), all three measured anisotropies have different temperature dependencies below 50~K. From our ESR results, we know that this can not be attributed to the temperature change of the $\bm{g}$ tensor, so we need to consider how anisotropic exchange interactions influence measured susceptibility anisotropy. In the case of the temperature independent $\bm{g}$ tensor we can write
\begin{equation}\label{eq:anisotropy}
\chi_x(T) - \chi_y(T)\propto g_x^2\; f_x(T) - g_y^2\; f_y(T).
\end{equation}
Functions $f_{x(y)}$ represent temperature dependence of susceptibilities measured along the axis $x(y)$. When interactions between spins are isotropic we have $f_x(T)=f_y(T)=f(T)$ which leads us back to expression \eref{eq:deltachiiso}. Anisotropic exchange interactions introduce different temperature dependencies of susceptibilities measured along different axes. This effect is observed both theoretically and experimentally in several low-dimensional $S=1/2$ systems. In $S=1/2$ 1D Heisenberg antiferromagnets DMI introduces anisotropic susceptibility enhancement.\cite{Affleck-1999,*Affleck-2000, Feyerherm-2000, Umegaki-2009, HerakPRB-2011} In kagome $S=1/2$ antiferromagnet both symmetric and antisymmetric anisotropic exchange strongly influence susceptibility anisotropy.\cite{Rigol-2007PRL, Rigol-2007, Han-2012} While the effect of anisotropic exchange on susceptibility might be weak if anisotropic exchange is weak, it is obvious that direct measurement of susceptibility anisotropy $\chi_x - \chi_y$ allowed by torque represents a much more sensitive method for detection of this effect, but only in combination with ESR $g$ factor measurements. In \cso we observe strongest disagreement between temperature dependences of susceptibility anisotropies measured in different planes in the temperature range where the $\bm{g}$ tensor is temperature independent (see \fref{fig4}). This suggests that anisotropic exchange needs to be introduced in the Hamiltonian describing this system. Further support to this claim is given by the temperature and angular dependence of the ESR linewidth.\\
\indent  Our estimation of dipole-dipole, anisotropic Zeeman and hyperfine coupling contributions to the linewidth show that the observed linewidth cannot be explained by these contributions only. The observed change of the direction of linewidth extrema with temperature also corroborates the presence of several competing contributions. All this strongly supports the presence of anisotropic exchange, both symmetric and antisymmetric. In \cso antisymmetric DMI is allowed by symmetry between spins from Cu1 and Cu2, but not between Cu1 and Cu1. From structural considerations we expect the interaction $J_{12}$ to be comparable in magnitude to $J_{11}$, although possibly smaller, so if the DMI affects the susceptibility anisotropy the effect could be observable at temperatures $T< J_{12}/k_B$. \\
\begin{figure}[t]
	\centering
		\includegraphics[width=0.80\linewidth]{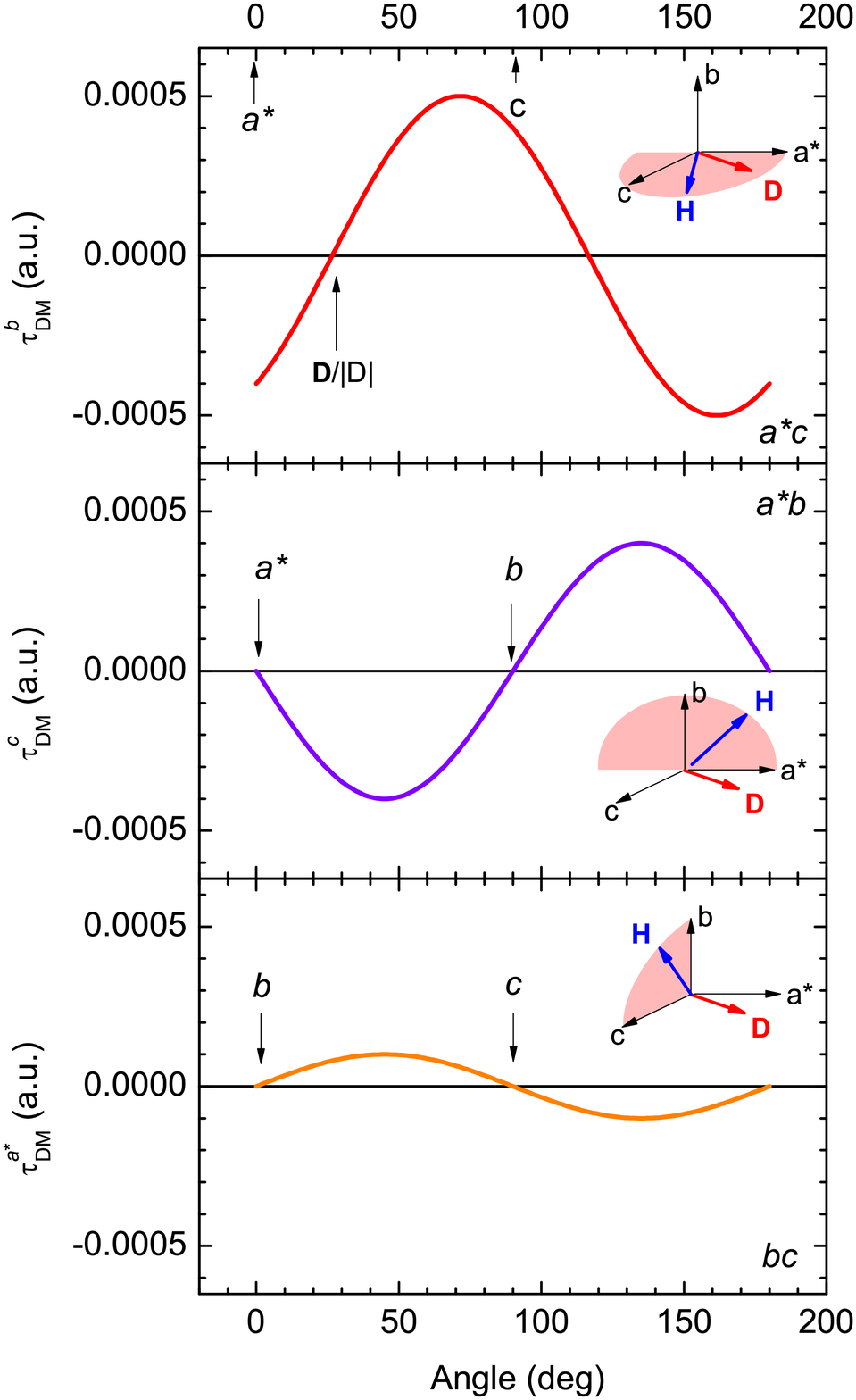}
	\caption{(Color online) Angular dependence of torque $\boldsymbol{\tau}_{DM} \propto \bm{m}_u^{DM} \times \bm{H}$ resulting from uniform magnetization \eref{eq:muniform} induced by DMI. For calculation $\bm{D}={(D_a^*,0,D_c)}$ and $D_c=0.5 D_{a^*}$ and isotropic $g=2$ was used. Scale on $y$ axis is equal for all panels. Direction $\bm{D}/|D|$ is the angle at which $\bm{H} \parallel \bm{D}$. Magnetic field rotates in plane spanned by axes shown in each panel. For each plane only the component of torque perpendicular to that plane is shown, as would be obtained from measurement.}
	\label{fig11}
\end{figure}
\indent Finally, we discuss the observed large rotation of magnetic axes below $\approx 50$~K where the $\bm{g}$ tensor is temperature -- independent. The observed rotation of magnetic axes at temperatures where the $\bm{g}$ tensor is temperature independent might also be a consequence of the anisotropic exchange interaction, namely, DMI. Symmetric anisotropic exchange tensor is diagonal in the same coordinate system in which the $\bm{g}$ tensor is diagonal. The DMI interaction
\begin{equation}\label{eq:DMI}
	\mathcal{H}_{\textup{DM}}= \bm{D} \cdot (\bm{S}_1 \times \bm{S}_2)
\end{equation}
 has vector $\bm{D}$ defined by crystal symmetry.\cite{Moriya-1960,*Moriya-1960PR} Inspection of Cu1 -- Cu2 connection suggests there are no restrictions on the direction of vector $\bm{D}$ in \csos.\cite{Moriya-1960,*Moriya-1960PR} We now try to understand how this affects macroscopic anisotropy measured by torque magnetometry. Torque is influenced by the component of total magnetization perpendicular to magnetic field. When only the $g$-factor anisotropy is present, magnetic axes are defined by the principal axes of the $\bm{g}$ tensor. When magnetic field is applied in the direction of the magnetic axes, induced magnetization $\bm{m}_u$ is parallel to the field and the resulting torque is zero. The rotation of the principal axes of the $\bm{g}$ tensor is thus reflected in the shift of the zeros of the sine curves of torque. This is what we observe in our measurements at $T\gtrsim 50$~K (see \fref{fig3}). Relatively recently it was shown that for antiferromagnetic spin $S=1/2$ dimer existence of the DMI between the spins on dimer induces a uniform magnetization \cite{Miyahara-2007} 
\begin{equation}\label{eq:muniform}
	\bm{m}_u^{\textup{DM}} \propto (\bm{D}\times \bm{H})\times\bm{D},
\end{equation}
where $\bm{D}$ is the DM vector, and $\bm{H}$ applied magnetic field. Equation \eref{eq:muniform} is obtained under assumption of isotropic $\bm{g}$. From \eqref{eq:muniform} we see that the vector $\bm{D}$ defines the magnetic axes: induced magnetization $\bm{m}_u^{\textup{DM}}$ is collinear to $\bm{H}$ when $\bm{H}\perp\bm{D}$ and $\bm{m}_u^{\textup{DM}}=0$ when $\bm{H}\| \bm{D}$, the torque being zero in all three cases. This is sketched in \fref{fig11} for a case when $\bm{D}={(D_a^*,0,D_c)}$ and $D_c=0.5 D_{a^*}$, where $a^*$, $b$, and $c$ represent three mutually perpendicular axes and the $\bm{g}$ tensor is assumed to be isotropic. We see that DMI contributes to torque in all three planes, but with different magnitudes. Magnetic axes defined by $\bm{g}$ tensor and by the DMI do not need to coincide. At high temperatures $T\geq J/k_B$ we expect that only the anisotropy of the $\bm{g}$ tensor is relevant. In case of antiferromagnetic dimer with two spins $S=1/2$ the magnetization $\bm{m}_u$ becomes vanishingly small at low temperatures (this is also true for AFM tetramer) while we expect $\bm{m}_u^{\textup{DM}}$ to increase as the temperature decreases. It is thus possible that at low temperature, $\bm{m}_u^{\textup{DM}}$ takes over inducing both the ''extra'' temperature dependence of susceptibility anisotropy and the rotation of the magnetic axes. Theoretical investigations of a spin tetramer system with anisotropic $\bm{g}$ tensor and the DMI should help clarify this. In Ref. \onlinecite{Miyahara-2007}, it was also shown that DMI induces large staggered magnetization $\bm{m}_s^{\textup{DM}} \propto \bm{D} \times \bm{H}$.  Further experimental confirmation of the presence of DMI in \cso could be obtained from NMR measurements which are a perfect local probe for detecting staggered magnetization.\\
\indent Finally, let us remark that the observed large rotation of macroscopic magnetic axes at low temperatures continues even in the magnetically ordered state (see bottom panel of \fref{fig5}). This is highly unusual and might be an evidence of a strong spin-lattice coupling in this system. 
\section{Conclusion}\label{sec:concl}
\indent Combining highly sensitive low-field torque magnetometry as a probe of macroscopic magnetic anisotropy and ESR measurements as a probe of microscopic anisotropy, we have shown that the previously detected rotation of macroscopic magnetic axes with temperature in \cso is a result of the rotation of the $\bm{g}$ tensor only at temperatures $T\gtrsim 50$~K. Disagreement of $<\chi>$ with the susceptibility of the proposed tetramer model \eref{eq:Hamiltonian}\cite{Emori-1975} is not a consequence of the temperature dependence of the $\bm{g}$ tensor since $<g>$ is temperature independent. Further theoretical and experimental investigations should clarify whether spin tetramer model applies to \csos. Temperature dependence of the measured $\bm{g}$ tensor can be explained by model of strongly coupled dimer within tetramer, however we point out some difficulties, namely the assumption that $J_{12} \ll J_{11}$ while structural considerations suggest they should be of comparable magnitudes. \\
\indent Estimated values of dipole-dipole interaction, the anisotropic Zeeman interaction and hyperfine coupling to the ESR linewidth are too small to explain the observed linewidth in \csos. This strongly suggests the presence of anisotropic exchange interaction in the Hamiltonian of \csos.\\
\indent Below 50~K $\bm{g}$ tensor becomes temperature -- independent, while very large rotation of macroscopic magnetic axes is observed in torque. Temperature dependencies of the susceptibility anisotropies in that temperature range differ substantially in three different crystallographic planes, and we argue that this is a result of the anisotropic exchange interaction, most likely DMI which is allowed by symmetry. We also propose that the same interaction is a reason behind the observed large rotation of magnetic axes in that temperature range. A small sharp kink at $T^*\approx 26$~K is detected  in susceptibility anisotropy in $(b, \; [1 0 1]^*)$ plane. The origin of this kink remains unknown.\\
\indent In the magnetically ordered state \cso is an antiferromagnet with easy axes along $\approx <\bar{1} 0 1>^*$ direction. Possible canting of spins cannot be detected by our bulk method and needs to be investigated by experiments which can detect orientations of individual spins such as neutron diffraction and muon spin relaxation. There is no net ferromagnetic moment in the ordered state. Rotation of magnetic axes observed below $T_N$ suggests strong spin-lattice coupling in this system.\\
\indent Experimental results presented here show that \cso is an interesting system for both theoretical and experimental study of the influence of small anisotropic exchange on ground state and excitations of quasi -- zero -- dimensional $S=1/2$ antiferromagnets.
\begin{acknowledgments}
M. H. is grateful to I. \v{Z}ivkovi\'{c} for critical reading of the manuscript and useful suggestions for its improvement. M. H. and A. G. \v{C} are grateful to N. Maltar Strme\v{c}ki and D. Cari\'{c} for technical support during ESR measurements. This work was supported by the resources of the Croatian Ministry of Science, Education and Sports under Grants No. 035-0352843-2846, No. 098-0982915-2939, and No. 035-0352843-2844 and the Croatian Science Foundation contract No. 02.05/33.
\end{acknowledgments}

\appendix
\section{\texorpdfstring{Calculation of the temperature dependent $\bm{g}$ tensor in \cso}{Calculation of the temperature dependent g tensor in SeCuO3}}\label{sec:app1}
\indent In the previous paper\cite{Zivkovic-2012}, it was suggested that the temperature dependence of the $\bm{g}$ tensor is a result of two different contributions to the total $\bm{g}$ tensor which have different temperature dependencies. It was shown that, in principle, it is possible to model temperature dependence of measured $\bm{g}$-tensor components under the assumption that $J_{12}=0$ and $J_{11}=290$~K, thus allowing two central spins from tetramer to form antiferromagnetic dimer which reduces their contribution to the total intensity as the temperature decreases. Here, we exploit this model using our measured $\bm{g}$ tensor and compare it to the temperature and angular dependencies obtained from experiment.\\
\indent Due to the morphology of the sample angular dependence of $\bm{g}$ tensor was measured in $(\xi, b, \eta)$ coordinate system where $\xi=[\bar{1} 0 1]$ and $\eta=[1 0 1]^* $. Total $\bm{g\:g^T}$ tensor can then be expressed as\cite{Weil}
\begin{align}\label{eq:ggT}
	\bm{g\:g^T} &= (\bm{g\:g^T})_{\xi\xi} \sin^2\theta \cos^2\varphi + (\bm{g\:g^T})_{b b}\sin^2 \theta \sin^2 \varphi  \\ \nonumber
	&+ (\bm{g\:g^T})_{\eta \eta} \cos^2 \theta + 2(\bm{g\:g^T})_{\xi b}\sin^2 \theta \sin \varphi \cos \varphi  \\ \nonumber
	&+ 2 (\bm{g\:g^T})_{\xi \eta} \cos \theta \sin \theta \cos \varphi +(\bm{g\:g^T})_{b \eta}\sin\theta \cos \theta \sin \varphi.
\end{align}
Angular dependence in the $ac$ plane shown in \fref{fig7} can then be fitted to \eref{eq:ggT} by setting $\varphi=0\deg$ and in the $(b, [101]^*)$ plane shown in \fref{fig8} by setting $\varphi=90\deg$  in \eref{eq:ggT}. In principle, to obtain all components of the $\bm{g\:g^T}$ tensor, it is necessary to perform measurements in three different crystallographic planes. However, from our previous results\cite{Zivkovic-2012} and from symmetry we know that only the off-diagonal element in the $ac$ plane is different from zero, so we set $(\bm{g\:g^T})_{\xi b}=0$ and $(\bm{g\:g^T})_{b \eta}=0$. Thus, our $(\bm{g\:g^T})$ tensor in the $(\xi, b, \eta)$ coordinate system reads as
\begin{equation}\label{eq:ggTxibeta}
	\bm{g\:g^T} =  \begin{bmatrix}
	(\bm{g\:g^T})_{\xi \xi} & 0 & (\bm{g\:g^T})_{\xi \eta}\\
	0 & (\bm{g\:g^T})_{b b} & 0 \\
	(\bm{g\:g^T})_{\xi \eta} & 0 & (\bm{g\:g^T})_{\eta \eta}
\end{bmatrix}.
\end{equation}
At temperatures 120 and 80~K, we use the average value of $(\bm{g\:g^T})_{\eta \eta}$ and $(\bm{g\:g^T})_{bb}$ obtained from fits to two angular dependencies in two different planes. At $T=15$~K, 40~K, and room temperature, we use $(\bm{g\:g^T})_{bb}$ obtained from temperature dependence of $g_b$ (\fref{fig9}). We diagonalize the obtained $\bm{g\:g^T}$ tensor, take a square root to obtain the diagonalized $\bm{g}$ tensor in the $(\xi, b, \eta)$ coordinate system and then perform a rotation to the $(a^*, b, c)$ coordinate system to obtain
\begin{equation}\label{eq:gasbc}
	\bm{g} =  \begin{bmatrix}
	\bm{g}_{a^* a^*} & 0 & \bm{g}_{a^* c}\\
	0 & \bm{g}_{b b} & 0 \\
	\bm{g}_{a^* c} & 0 & \bm{g}_{c c}
\end{bmatrix}.
\end{equation}
 The measured components of tensor $\bm{g}$ at different temperatures are given in Table \ref{tab:gcomponents}. \\
\begin{table}[tb]
\centering
		\begin{tabular*}{\linewidth}{@{\extracolsep{\fill}}c | c c c c }
		\hline \hline
			T~(K) & $g_{a^* a^*}$ & $g_{b b}$	& $g_{c c}$ & $g_{a^* c}$ \\	
				\hline
			  293 & 2.2056 & 2.2394  & 2.1258 & 0.0344 \\
				120 & 2.1729 & 2.2967 & 2.0907 & 0.0364 \\
				80 & 2.1536 & 2.3223 & 2.0881 & 0.0382 \\
				40 & 2.1409 & 2.3367 & 2.0895 & 0.0371 \\ 
				15 & 2.1379 & 2.3350 & 2.0896 & 0.0379\\ 
			\hline \hline	
				\end{tabular*}
\caption{ $\bm{g}$ tensor components in the $(a^*, b, c)$ coordinate system measured at different temperatures.}
\label{tab:gcomponents}
\end{table}
\begin{figure*}[t]
\centering
		\includegraphics[width=\textwidth]{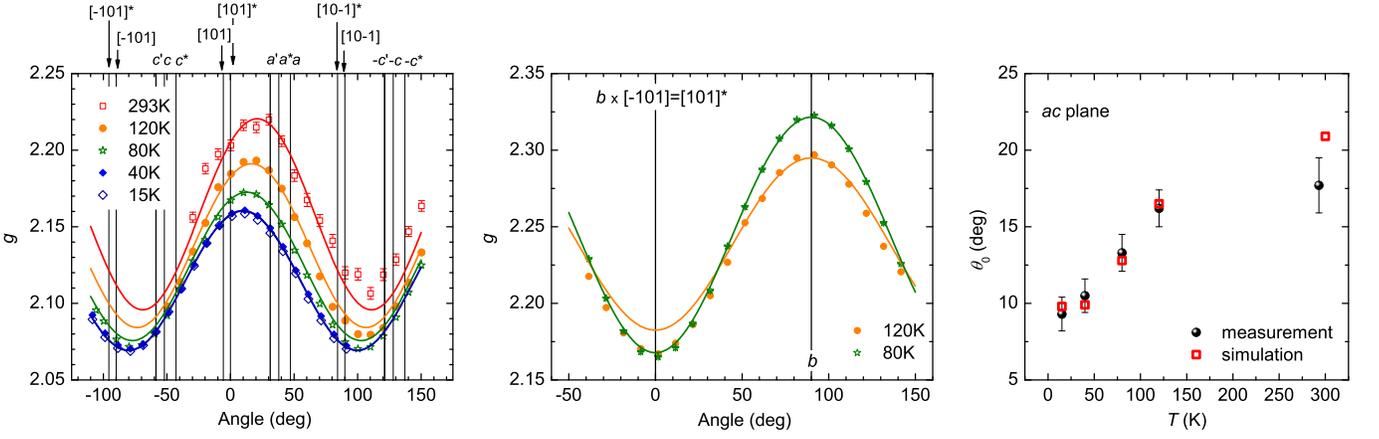}
	\caption{(Color online) Left and middle panel: Comparison of calculated angular dependence of $g$ factor in $ac$ plane (left panel) and $(b,[101]^*)$ plane (middle panel) with the measured one at different temperatures. Right panel: Temperature dependence of the phase $\theta_0$ of the $\bm{g}$ tensor in the $ac$ plane obtained from the measurements compared to the results of calculations presented in Appendix \ref{sec:app1}. Presented calculations used experimentally obtained $\bm{g}$ tensor components. }
	\label{fig12}
\end{figure*}
\indent Symmetry restrictions will help us to approximate the individual tensors $\bm{g}_1$ and $\bm{g}_2$ belonging to crystallographically inequivalent Cu1 and Cu2. There are two magnetically inequivalent tetramers, as can be seen in \fref{fig1}. We can term them A and B. Spins $\bm{S}_2$ and $\bm{S}_3$ from Cu1 on one tetramer have equivalent $\bm{g}$ tensors because there is an inversion center between two Cu1 ions carrying those spins. The same is true for $\bm{S}_1$ and $\bm{S}_4$ from the same tetramer. Thus, there are four magnetically inequivalent $\bm{g}$ tensors, two belonging to Cu1 and two to Cu2. We term them $\bm{g}_{1A}$, $\bm{g}_{1B}$, $\bm{g}_{2A}$, and $\bm{g}_{2B}$, where subscript 1(2) denotes Cu1 (Cu2) and subscript A (B) denotes tetramer A (B). Symmetry requirements further impose $g_{I A}^{a^* a^*}=g_{I B}^{a^* a^*}$, $g_{I A}^{b b}=g_{I B}^{b b}$, $g_{I A}^{c c}=g_{I B}^{c c}$, $g_{I A}^{a^* c}= g_{I B}^{a^* c}$, $g_{I A}^{a^* b}= - g_{I B}^{a^* b}$, $g_{I A}^{b c}= - g_{I B}^{b c}$, where $I = 1,2$. It is obvious that we cannot obtain real $\bm{g}$ tensors of each magnetically inequivalent Cu, but we can get approximate tensors $\bm{g}_1=(\bm{g}_{1A}+ \bm{g}_{1B})/2$ and $\bm{g}_2=(\bm{g}_{2A}+ \bm{g}_{2B})/2$. \\
\indent Following Ref. \onlinecite{Zivkovic-2012} we now assume that the central Cu1-Cu1 pair forms an antiferromagnetic dimer with interaction $J_{11}$ between the two spins and that the probability of finding a dimer in a triplet state is given by\cite{Zivkovic-2012} 
\begin{equation}\label{eq:dimer}
  \alpha(T) = \dfrac{3 \: e^{-J_{11}/k_B T}}{1+3\:e^{-J_{11}/k_B T}}.
\end{equation}
We can then write for the temperature-dependent $\bm{g}$ tensor
\begin{equation}\label{eq:gtotalT}
	\bm{g}(T) = \dfrac{\alpha}{\alpha+\beta} \; \bm{g}_{1} 
	+ \dfrac{\beta}{\alpha+\beta} \;\bm{g}_{2}
\end{equation}
where $\alpha=\alpha(T)/\alpha_{max}$, $\alpha_{max}=3/4$, $\beta=1$. Assuming that the symmetry does not change with temperature we calculated a value of $\bm{g}$ tensor in the direction of magnetic field $\bm{n}=\bm{B}/B$, $g_n$, using $ g_n = \sqrt{\bm{n^T}\cdot \bm{g\:g^T}\cdot \bm{n}}$,\cite{Weil} where $\bm{g\:g^T}$ at certain temperature was determined using \eref{eq:gtotalT} with $J_{11}=290$~K. For large $J_{11}$ only the contribution from $\bm{g}_2$ should remain at low temperatures. We thus set $\bm{g}_2 \approx \bm{g}(15~K)$. The measured data are best simulated by setting $\bm{g}_1^{a^* a^*}= 2.3082$, $\bm{g}_1^{b b}= 2.1351$, $\bm{g}_1^{c c}= 2.1304$, $\bm{g}_1^{a^* c}= 0.0309$ and $\bm{g}_2^{a^* a^*}= 2.1399$, $\bm{g}_2^{b b}= 2.3400$, $\bm{g}_2^{c c}= 2.0896$, $\bm{g}_2^{a^* c}= 0.0379$. In \fref{fig9} we compare calculated temperature dependence of components $g_{a'}$, $g_b$, and $g_{c'}$ with the measured ones. The agreement is quite satisfactory. In \fref{fig12} it can be observed that the temperature change of angular dependencies is also captured along with the change of $\theta_0$ - the shift of the $g$ factor maximum with temperature in the $ac$ plane.\\ 
\begin{figure}[b]
	\centering
		\includegraphics[width=0.70\linewidth]{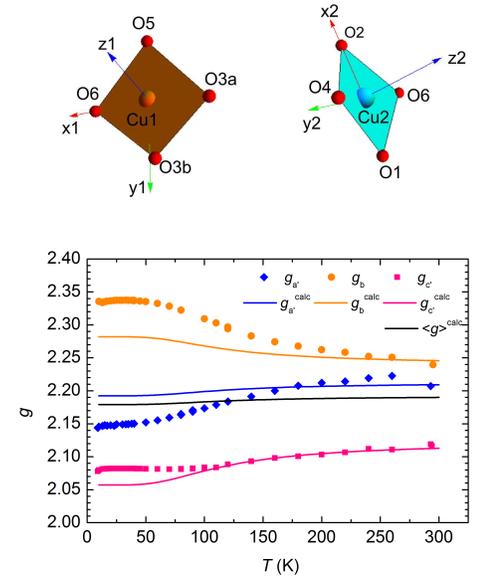}
		\caption{(Color online) Top: Two crystallographically inequivalent CuO$_4$ distorted squares in \cso. Principal axes of $\bm{g}$ tensors are also shown. Bottom: Temperature dependence of $\bm{g}$ tensor components calculated using principal axes shown above (see text).}
	\label{fig13}
\end{figure}
\indent Although the above-presented simulation describes the observed behavior of the $\bm{g}$ tensor very well, it has a disadvantage that from the measured values of the $\bm{g}$ tensor we can not obtain the tensor components $g_{a^*b}$ and $g_{bc}$ which, from crystal symmetry, we expect to be significant. Thus, we can not determine the principal $\bm{g}$ tensors of Cu1 and Cu2. To do that, we try a different approach: using purely structural considerations we construct the principal axes of the tensors $\bm{g}_1$ and $\bm{g}_2$. Then, we determine the principal components of each tensor in such a way to obtain the measured values at room temperature given in Table \ref{tab:gcomponents}.\\
\indent The CuO$_6$ octahedra in \cso are distorted, which means there could be some small admixture of the $d_{3z^2-r^2}$ orbital in the ground-state orbital. In \fref{fig13}, we show CuO$_4$ distorted squares where we assume the $d_{x^2-y^2}$ orbital is located. We then assign the principal values of Cu1 and Cu2 as follows. In Cu1O$_4$ atoms, O(3a) and O(3b) belong to two Cu1 atoms which form a dimer. Together with O(6) they form a plane. We assume that this plane is an equatorial plane where principal $x$ and $y$ axes lie. We assign principal axis $x_1$ along the Cu1-O(6) direction, $y_1$ is then close to the Cu1-O(3b) direction, and $z_1$ is perpendicular to both. Cu2O$_4$ is even more distorted. We define $x_2$ along Cu2-O(2), $y_2$ is approximately along Cu2-O(4), and $z_2$ perpendicular to both. Similar configuration of CuO$_4$ and construction of the  principal axes is found and carried out for CuTe$_2$O$_5$.\cite{Deisenhofer-2006} Thus assigned principal axes are shown in the top of \fref{fig13}. We then look for values of principal tensors which best describe our data at room temperature. These are $\bm{g}_1=(2.1,\;2.19,\;2.33)$ and $\bm{g}_2=(2.035,\; 2.185,\; 2.31)$. We express each magnetically inequivalent tensor in the $(a^*,b,c)$ coordinate system and apply \eqref{eq:gtotalT} to calculate the temperature dependence of different $\bm{g}$-tensor components. The result is shown in the bottom of \fref{fig13}. Although the general behavior is captured, the obtained values are very different from the measured ones. We have also attempted other reasonable assignments of principal axes, but it turns out that it is not possible to find two tensors $\bm{g}_1$ and $\bm{g}_2$ with values of their components which are physically acceptable for Cu$^{2+}$ spin $S=1/2$ which completely describe the temperature dependence of the measured data using \eqref{eq:gtotalT}. This reflects the fact that the distortion of the CuO$_6$ might be more significant than our simplified approach takes into account, but also that the assumption of \eqref{eq:gtotalT} that $J_{12}\ll J_{11}$ is incorrect. 
\newpage
 %Just because of unusual number of tables stacked at end
\bibliographystyle{apsrev4-1}% Produces the bibliography via BibTeX.
\bibliography{Ref}
%
%\newpage
\end{document}